\documentstyle[citation,epsfig]{article}

\font\tenrm=cmr10
\font\tenit=cmti10
\font\elevenbf=cmbx10 scaled\magstep 1
\font\elevenrm=cmr10 scaled\magstep 1
\font\elevenit=cmti10 scaled\magstep 1

\renewenvironment{thebibliography}[1]
 { \elevenrm
   \begin{list}{\arabic{enumi}.}
    {\usecounter{enumi} \setlength{\parsep}{0pt}
     \setlength{\itemsep}{3pt} \settowidth{\labelwidth}{#1.}
     \sloppy
    }}{\end{list}}

\begin{document}
\begin{center}
\rightline{\vbox{ 
\halign{&#\hfil\cr
&FERMILAB-CONF-93/348-T\cr
&MAD/PH/802\cr
&NUHEP-TH-93-28\cr
&OCIP/C-93-14\cr}}} 
\vglue 0.6cm
{
 {\elevenbf        \vglue 10pt
               CHARMONIUM PRODUCTION VIA FRAGMENTATION\\
               \vglue 3pt
               AT {\em {\elevenbf p\={p}}} COLLIDERS 
\\}
\vglue 1.0cm
{\tenrm Michael A. Doncheski\footnote{Address after 15 August 1993 - Physics
Department, Carleton University, Ottawa, Ontario K1S 5B6, Canada.}\\}
\baselineskip=13pt
{\tenit Physics Department , University of Wisconsin, Madison, Wisconsin 
53706, USA \\}
\vglue 0.3cm
{\tenrm Sean Fleming\\}
{\tenit Department of Physics, Northwestern University, Evanston, Illinois 
60208, USA\\}
\vglue 0.3cm
{\tenrm Michelangelo L. Mangano\\}
{\tenit INFN, Scuola Normale Superiore and Dipartimento di Fisica, Pisa, 
Italy \\}}

\vglue 0.8cm
{\tenrm ABSTRACT}

\end{center}

\vglue 0.3cm
{\rightskip=3pc
 \leftskip=3pc
 \tenrm\baselineskip=12pt
 \noindent
We present the preliminary results of a calculation of the 
fragmentation contribution to charmonium production at large transverse 
momentum in $p\bar{p}$ colliders. The fragmentation of gluons and charm quarks 
is the dominant direct production mechanism for sufficiently large $p_{\perp}$.
We find that for both $J/\psi$ and $\psi '$ production fragmentation dominates 
over the conventional gluon-gluon fusion mechanism for $p_{\perp}$ greater than
about 6 GeV.}
\\
\vglue 0.3cm
{\rightskip=3pc
 \leftskip=3pc
 \tenrm\baselineskip=12pt
 \noindent
Presented at the Workshop on Physics at Current Accelerator and the
Supercollider, Argonne National Laboratory, June 1993.
}

\vglue 0.6cm
\baselineskip=14pt
\elevenrm
The study of charmonium production in high energy hadronic collisions provides 
a fundamental testing ground for perturbative quantum chromodynamics (QCD). 
Through the comparison of experimental data and theory it is possible 
to check if the basic ideas used in these calculations are correct. The study 
of $J/ \psi$ production is of particular importance because the decay 
$B \rightarrow \psi + X$ plays a crucial role in the measurements of $b$ quark 
production and the studies of CP violation in the B meson system.  In order 
to properly understand these phenomena, it is necessary to understand the
background due to direct production of $\chi$ and $\psi$. 

Previous calculations of direct charmonium production at large $p_{\perp}$ in 
$p \bar{p}$ collisions have assumed that the dominant contribution to the cross
section comes from those leading order diagrams with gluon-gluon fusion into a 
charmonium state and a recoiling gluon \cite{br}. These calculations do not 
reproduce all aspects of the available data \cite{mlm}, which suggests that
there are other important production mechanisms. Recently it was pointed out 
that there are fragmentation contributions to charmonium  production that come 
{}from higher order in the strong coupling constant $\alpha_{s}$, but 
eventually dominate because of a softer $p_{\perp}$ dependence \cite{by}. In 
these proceedings we present preliminary results of a project to calculate the 
fragmentation contributions to charmonium production. Our completed 
calculations are the fragmentation contributions to the direct production of 
$\psi$ and $\psi '$. The $\psi '$ results can be added to the contributions 
{}from gluon-gluon fusion and the decay $B \rightarrow \psi ' + X$ and 
compared with CDF data on inclusive $\psi '$ production. The inclusive 
production of the $\psi$ is complicated by the fact that the $\chi$ states 
decay via $\chi \rightarrow \psi + \gamma$, so it is important that direct 
$\chi$ production be included when comparing with data on inclusive $\psi$
production. The fragmentation contribution to $\chi$ production needs to be
calculated before this comparison can be made. We can, however, compare the 
fragmentation contribution to direct $\psi$ production to the gluon-gluon 
fusion contribution and determine above what $p_{\perp}$ the fragmentation 
contribution dominates.
 
Fragmentation is the process in which a high $p_{\perp}$ parton is created and
subsequently decays into a hadron. There is a rigorous factorization theorem 
of perturbative QCD that applies to the inclusive production of such hadrons 
in $e^{+} e^{-}$ annhilation \cite{cs}. It states that the differential
cross section $d \sigma$ for producing a hadron with large $p_{\perp}$ factors
into differential cross sections $d \hat{\sigma}$ for producing partons with
large transverse momentum and fragmentation functions $D(z)$. The fragmentation 
function gives the probability for the splitting of the parton into the hadron
with momentum fraction $z$, 
and is universal in the sense that it is independent of the process that
produces the fragmenting parton. It has recently been shown that the
fragmentation functions $D(z)$ for heavy quarkonium states can be calculated
using perturbative QCD \cite{by}. Fragmentation functions for several of these
states have been calculated explicitly \cite{by,bcy,flsw,c}.

The factorization theorem for $e^{+} e^{-}$ annihilation can be generalized to 
the inclusive production of a hadron at large $p_{\perp}$ in $p \bar{p}$ 
collisions. For sufficiently large $p_{\perp}$, the differential cross section 
should have a factorized form in terms of parton distribution functions, hard
scattering cross sections, and fragmentation functions. Taking $\psi$ 
production to be specific, the differential cross section can be written
\begin{equation}
d\sigma(p\bar{p} \rightarrow \psi (p_{\perp},y)+X)= 
\sum_{i} \int^{1}_{0} \! dz \; d \sigma(p\bar{p} \rightarrow 
i(\frac{p_{\perp}}{z},y)+X, \mu) \; D_{i \rightarrow  \psi}(z, \mu),
\label{one}
\end{equation}
where the sum is over partons of type $i$, $z$ is the longitudinal momentum
fraction of the $\psi$ relative to the parton, and $y$ is the rapidity of the
$\psi$. $D_{i \rightarrow \psi}(z, \mu)$ is the fragmentation function, and 
$\mu$ is the factorization scale which cancels between the two factors. The 
fragmentation functions can be evolved to any scale via the Altarelli-Parisi 
evolution equations 
\begin{equation}
\mu \frac{\partial}{\partial \mu} D_{i \rightarrow \psi}(z, \mu) = 
\sum_{j} \int^{1}_{z} \frac{dy}{y} \; P_{i \rightarrow j}(z/y, \mu) 
\; D_{j \rightarrow \psi }(y, \mu),
\label{onehalf}
\end{equation}
where $P_{i \rightarrow j}(x, \mu)$ is the Alterelli-Parisi splitting function
for the splitting of a parton of type $i$ into a parton of type $j$ with a
longitudinal momentum fraction $x$. The dominant contributions to Eq. 
(\ref{one}) come from gluon fragmentation and charm quark fragmentation. At 
leading order in $\alpha_{s}$ the hard gluon or charm quark is produced by a 
hard $2 \rightarrow 2$ scattering process and the differential cross section 
on the right side of Eq. (\ref{one}), neglecting parton masses, is 
\begin{eqnarray}
\lefteqn{\frac{d \sigma}{d p^{2}_{\perp} dy}(p\bar{p} \rightarrow 
i(\frac{p_{\perp}}{z},y)+X) =  }
\nonumber
\\
& & \sum_{j k}\int^{1}_{0} \! dx_{1} \,
f_{p \rightarrow j}(x_{1}) \int^{1}_{0} \! dx_{2} \, 
f_{\bar{p} \rightarrow k}(x_{2}) 
\, \frac{1}{z} \, \delta \! \left( z-( \frac{e^{y}}{x_{1}} + 
\frac{e^{-y}}{x_{2}})
\frac{p_{\perp}}{\sqrt{s}} \right) \, 
\frac{d \hat{ \sigma}}{d \hat{t}},
\label{four}
\end{eqnarray}
where $x_{1}$ and $x_{2}$ are the longitudinal momentum fraction of partons
j and k relative to the proton and antiproton, $f_{p \rightarrow j}(x_{1})$ and 
$f_{\bar{p} \rightarrow k}(x_{2})$ are the corresponding parton distribution
functions, $s$ is the center of mass energy squared, and 
$d \hat{ \sigma} / d \hat{t}$ is the differential cross section. The 
differential cross sections $d \hat{\sigma} /d \hat{t}$ for the relevant 
subprocesses can be found in standard texts \cite{b}. For a given $y$, 
$p_{\perp}$, and $s$, the delta function in Eq. (\ref{four}) will translate 
into a limit on $x_{1}$ and $x_{2}$, where
\begin{equation}
0 < \frac{e^{-y}}{x_{2}} + \frac{e^{y}}{x_{1}} < \frac{\sqrt{s}}{p_{\perp}}.
\label{seven}
\end{equation}

The fragmentation functions $D_{g \rightarrow \psi}(z, \mu )$ and 
$D_{c \rightarrow \psi}(z, \mu )$ were calculated in Ref. 3 and 5 to leading
order in $\alpha_{s}$ at a scale $\mu$ on the order of $2 m_{c}$. We used
Eq. (\ref{onehalf}) to evolve them to the appropriate scale $\mu = p_{\perp}/z$ 
set by the transverse momentum of the parton. Only the splitting term 
$g \rightarrow g$ was used in the evolution of $D_{g \rightarrow \psi}$, and 
only the splitting term $c \rightarrow c$ was used in the evolution of 
$D_{c \rightarrow \psi}$. It has been pointed out that the $g \rightarrow c$ 
splitting term may be important in the evolution of the gluon fragmentation 
function \cite{flsw}. 

The evolved fragmentation functions and Eq. (\ref{four}) were used in 
Eq. (\ref{one}) to calculate the fragmentation contribution to direct $\psi$
production. The MRSD0 parton distribution set wes used along with a 
pseudorapidity cut of $| \eta | < 0.5$, scale $\mu = 2 \mu_{0}$ and a beam 
energy of 900 GeV. The result of the calculation is shown in figure~1. The 
gluon fragmentation and charm 
quark fragmentation contributions are graphed separately and compared with
the gluon-gluon fusion contribution. The cross over $p_{\perp}$ value above 
which fragmentation dominates is quite low. For gluon fragmentation the
cross over $p_{\perp}$ is about 8 GeV while for charm quark fragmentation it 
is about 6 GeV.  In spite of a suppression by two powers of $\alpha_{s}$, the 
fragmentation contribution dominates at large $p_{\perp}$ because the 
subprocess differential cross section $d \hat{\sigma}/d \hat{t}$ scales as
$1/p^{4}_{\perp}$ whereas for gluon-gluon fusion it scales as 
$1/p^{8}_{\perp}$. 
\begin{figure}
\begin{center}
\mbox{\epsfig{file=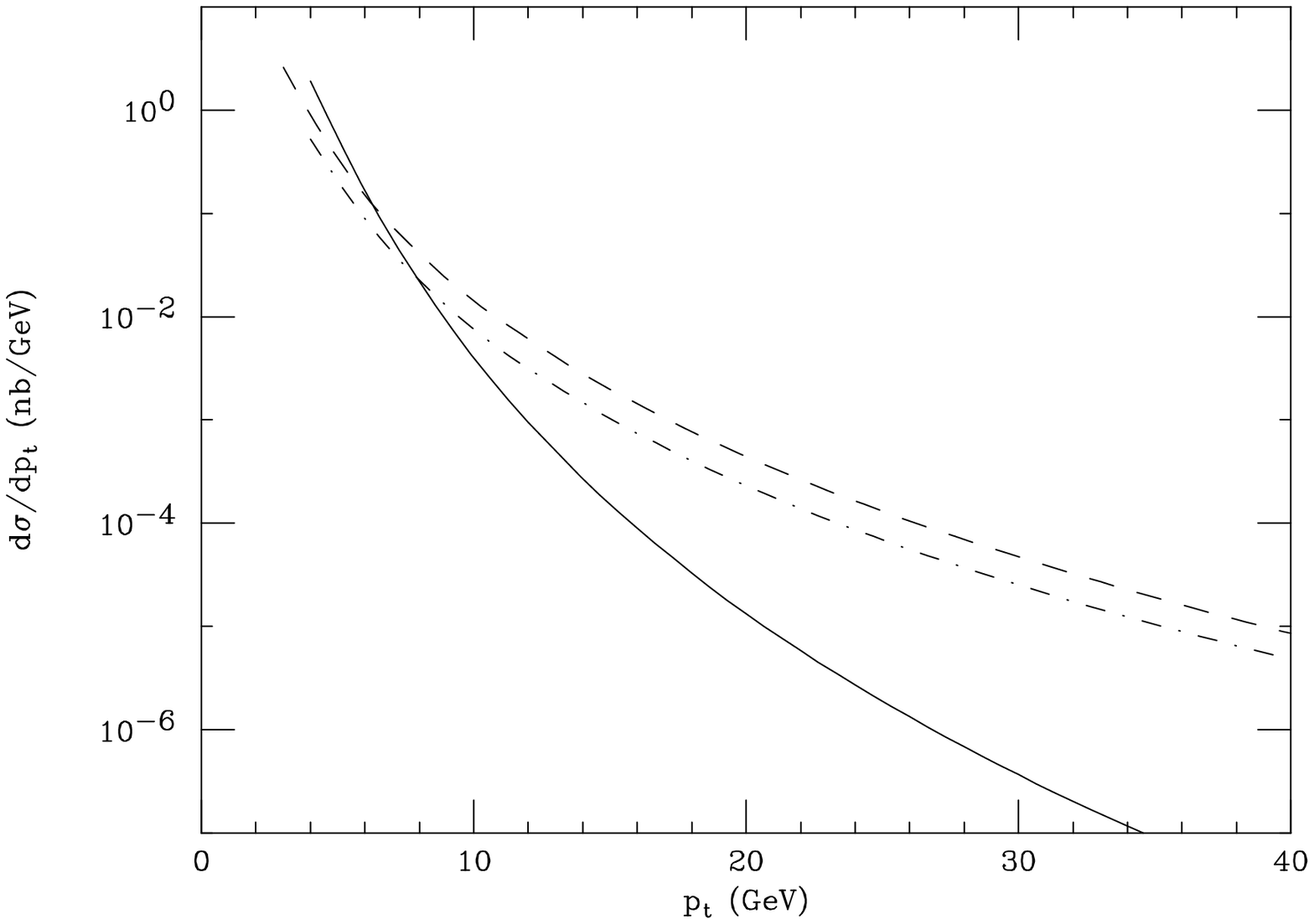,height=8.0cm}}
\end{center}
\caption{Direct $\psi$ production at the Tevatron. Gluon-gluon fusion with 
$\mu^{2}_{0} = M^{2}_{\psi} + p^{2}_{\perp}$ (solid), gluon fragmentation with 
$\mu_{0} =p^{gluon}_{\perp}$ (dotdash), charm quark fragmentation with 
$\mu_{0} =p^{charm}_{\perp}$ (dashes)}
\end{figure}

The $\psi'$ fragmentation contribution can be obtained from the $\psi$
fragmentation contribution simply by multiplying by the ratio of their
electronic widths. The results at Tevatron energies are shown in figure~2. 
The parton distribution set, the scale $\mu_{0}$ and the psuedorapidity cut are 
the same as those used in 
the $\psi$ calculation. The two fragmentation contributions are shown as well 
as the result of the gluon-gluon fusion calculation, the 
$B \rightarrow \psi ' + X$ calculation, and the CDF data \cite{cdf}. The
cross over points are the same as those in $\psi$ production. It is clear 
{}from the graph that the theoretical predictions do not properly reproduce the 
shape of the data, a sign that there may still be other production mechanisms 
that need to be taken into account. 
\begin{figure}
\begin{center}
\mbox{\epsfig{file=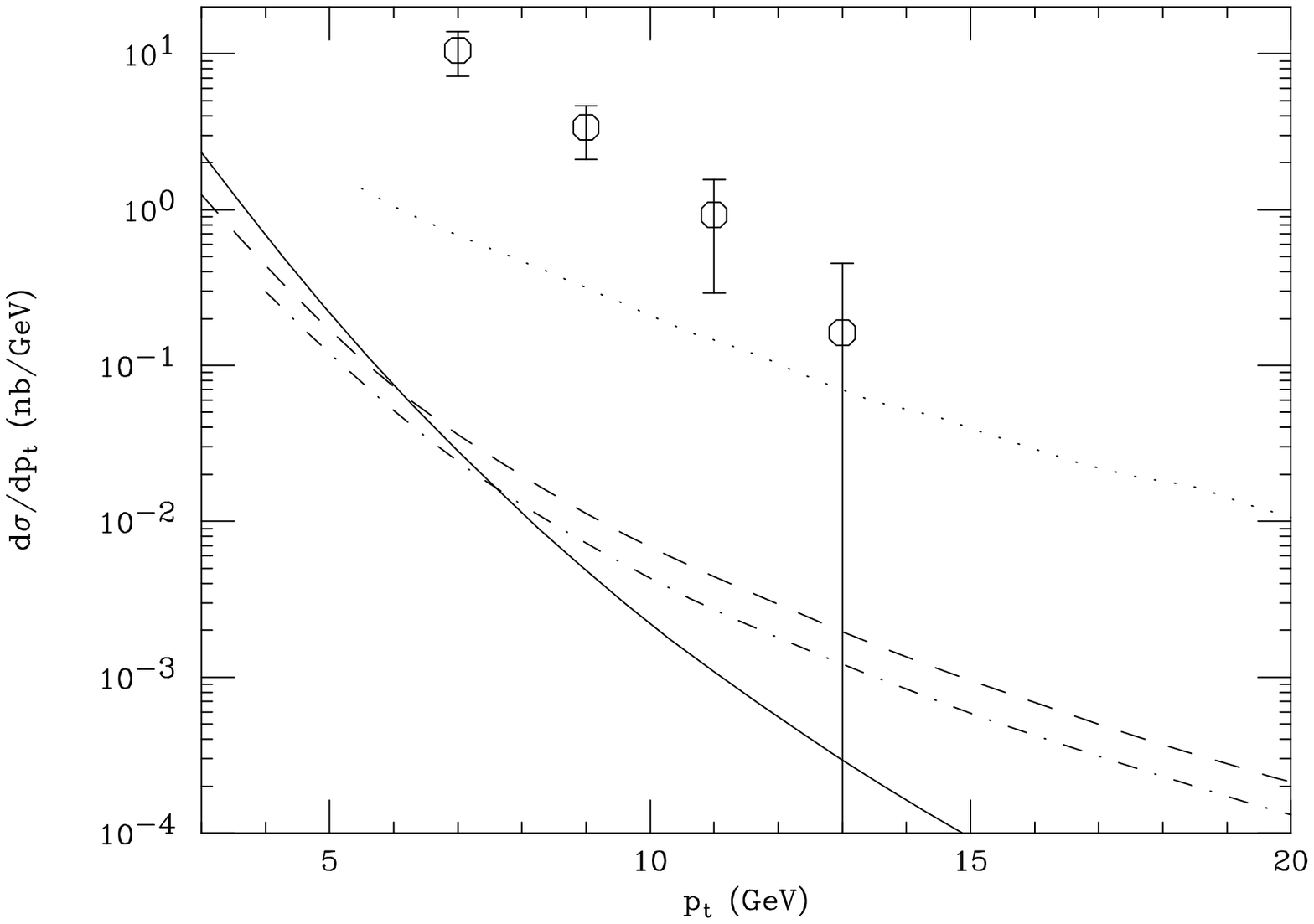,height=8.0cm}}
\end{center}
\caption{$\psi '$ production at the Tevatron. CDF data (O) compared to 
gluon-gluon fusion
with $\mu_{0}^{2} = M^{2}_{\psi '} + p^{2}_{\perp}$ (solid), gluon fragmentation
with $\mu_{0} =p^{gluon}_{\perp}$ (dotdash), charm quark fragmentation with 
$\mu_{0} = p^{charm}_{\perp}$ (dashes), and $B \rightarrow \psi ' + X$ (dots).}
\end{figure}

Some important comments about the results of the calculation need to be made. 
We have demonstrated that although the fragmentation contribution is of higher 
order in $\alpha_{s}$, it will dominate the lower order gluon-gluon fusion 
process for $p_{\perp}$ greater than about 6 GeV. This is small 
enough that the fragmentation contribution is relevant to the analysis of 
present collider experiments. At higher energies, such as those 
that may be reached at the LHC and SSC, the fragmentation contribution will be 
especially important because of the large range of $p_{\perp}$ that detectors 
will be able to probe. Another important point is that because of the 
fragmentation contribution the background to $B \rightarrow \psi + X$ from 
direct $\psi$ or $\chi$ production will not vanish at large $p_{\perp}$, 
contrary to some previous suggestions. The gluon-gluon fusion contributions 
fall away rapidly at large $p_{\perp}$ since they involve parton subprocesses 
for which $d \hat{\sigma}/d \hat{t}$ scales as $1/p^{8}_{\perp}$ for 
$\psi$'s or $1/p^{6}_{\perp}$ for $\chi$'s. However the fragmentation 
contribution involves parton subprocesses for which $d \hat{\sigma}/d \hat{t}$ 
scales as $1/p^{4}_{\perp}$ just like the cross section for $b$-quark 
production. A final point that needs to be made is that isolation cuts on the
$\psi$ can not eliminate the background to  $B \rightarrow \psi + X$ from 
direct $\psi$ production, contrary to previous suggestions. While the 
gluon-gluon fusion mechanism does produce isolated $\psi$'s and $\chi$'s, 
the fragmentation mechanism produces $\psi$'s and $\chi$'s inside gluon 
jets or charm quark jets.

Once again we want to stress that this is not a complete result for either 
$\psi$ or $\psi'$. In the case of $\psi$ production no comparison to data can
be made until the contribution from direct $\chi$ production is calculated 
completely, including direct $\chi$'s produced from fragmentation. In case of 
the $\psi'$, we have made a comparison with data and it suggests that there 
may still be other production mechanisms for charmonium that have not yet been 
identified. 

The work of MAD was supported, in part by the U. S. Department of Energy under
Contract No. DE-AC02-76ER00881, in part by the Texas National Research 
Laboratory Commission under Grant Nos. RGFY9273 and  RGFY93-221, in part by the 
University of Wisconsin Research Committee with funds granted by the Wisconsin 
Alumni Research Foundation and in part by the Natural Sciences and Engineering
Research Council of Canada. The work of SF was supported in part by the U.S. 
Department of Energy, Division of   High Energy Physics, under Grant 
DE-FG02-91-ER40684. The work of MLM was supported by INFN, Scuola Normale 
Superiore and Dipt. di Fisica, PISA, ITALY.

\end{document}